\documentclass[epj]{svjour}
\usepackage{graphics}
\usepackage{tensor}
\usepackage{amssymb}
\usepackage{feynmp}
\usepackage{psfrag}

\begin{document}
\begin{fmffile}{diagrams}

\title{Large scale shell model calculations for odd-odd $^{58-62}$Mn isotopes}
\author{P. C.~Srivastava$^{1,2,3}$ \thanks{e-mail: chandrapraveen1@gmail.com}
\and I.~Mehrotra$^1$}
\institute{ Nuclear Physics Group, Department of Physics, University of Allahabad-211002, India 
\and Grand Acc\'el\'erateur National d'Ions Lourds, CEA/DSM--CNRS/IN2P3, BP~55027, F-14076 Caen Cedex 5, France
\and  Physical Research Laboratory, Ahmedabad 380 009, India}
\date{\today}

\abstract{ Large scale shell model calculations have been carried out for odd-odd $^{58-62}$Mn isotopes in two different model spaces. First set of calculations have been carried out in full $\it{fp}$ shell valence space with two recently derived $\it{fp}$ shell interactions namely GXPF1A and KB3G treating $^{40}$Ca as core. The second set of calculations have been performed in ${fpg_{9/2}}$ valence space with the $fpg$ interaction treating $^{48}$Ca as core and imposing a truncation by allowing up to a total of six particle excitations from the 0f$_{7/2}$ orbital to the upper  $\it{fp}$ orbitals for protons and from the upper $\it{fp}$ orbitals to the 0g$_{9/2}$ orbital for neutron. For low-lying states in $^{58}$Mn, the KB3G and GXPF1A both predicts good results and for $^{60}$Mn, KB3G is much better than GXPF1A. For negative parity and high-spin positive parity states in both isotopes $fpg$  interaction is required. Experimental data on $^{62}$Mn is sparse and therefore it is not possible to make any definite conclusions. More experimental data on negative parity states is needed to ascertain the importance of 0g$_{9/2}$ and higher orbitals in neutron rich Mn isotopes.
}

%\pacs{21.60.Cs, 27.50.+e}  
\maketitle

\section{Introduction}
\label{s_intro}
 The region of the neutron-rich $\it{fp}$ shell nuclei between the magic numbers N=28 and N=50 has been at the focus of recent experimental and theoretical studies. The actual mechanism which causes the changes in nuclear structure as neutron number increases in a nuclear system is still an open question. Neutron rich $\it{fp}$ shell nuclei are also of special interest in astrophysics such as the electron capture rate in supernovae explosion. However the experimental data on neutron rich $\it{fp}$ shell nuclei is limited to those that can be populated by different experimental techniques. Recently energy levels of odd-odd  $^{58-62}$Mn isotopes have been reported in the literature. Appelbe {\it et al.} \cite{App00} have populated the high spin states of $^{58}$Mn by bombarding 40 MeV $^{13}$C  beam upon an enriched $^{48}$Ca target. The level scheme of $^{59}$Mn was reported by Liddick {\it et al.} \cite{Liddick05} following $\beta$-decay of $^{59}$Cr. The excited states of $^{60}$Mn were populated following the $\beta$-decay of $^{60}$Cr \cite{Liddick06}. Gaudefroy {\it et al.} \cite{Gau05} have studied  $\beta$ decay of $^{62}$Cr to $^{62}$Mn, but the location and ordering of the proposed 1$^+$ and 4$^+$ states in $^{62}$Mn are very puzzling. Valiente-Dob\'on  {\it et al.} \cite{Dob08} studied neutron-rich Mn isotopes from A=59-63, through multi-nucleon transfer reaction by bombarding a $^{238}$U target with 460 MeV $^{70}$Zn beam. The identification of $\gamma$-rays belonging to each nucleus was carried out with high precision by coupling the clover detector of Euroball (CLARA) to PRISMA   magnetic spectrometer. This experiment has provided data on the level structure of neutron rich Mn isotopes from A=59 to 63. Crawford {\it et al.} \cite{Cra09} have studied  the $\beta$ decay of  $^{61}$Cr for non-yrast excited states in the daughter nucleus $^{61}$Mn. More recently D. Steppenbeck {\it et al.} \cite{Ste10} have populated high-spin structure in neutron rich $^{57-60}$Mn isotopes with fusion-evaporation reactions induced by  $^{48}$Ca beams at 130 MeV on $^{13,14}$C targets at Argonne National Laboratory. Also for many low-lying levels tentative $J^\pi$ assignments have been made in this paper. Theoretically importance of the intruder 0g$_{9/2}$ orbital for  lighter $\it{fp}$ shell nuclei namely neutron rich Cr and Fe isotopes are recently reported in the literature \cite{Lurandi07,Kaneko08}.

 The paper is organized as follows: sect. 2 gives details of the calculation. Results and discussion are given in sect. 3. Finally in sect. 4 a summary is given.

\section{ Details of Calculation}

  In the present work we have performed large scale shell model calculations on neutron rich $^{58-62}$Mn isotopes in two different model spaces. The aim of the present work is to test (i) the recent experimental data on neutron rich Mn isotopes (ii) suitability of chosen valence space  (iii) effective interactions for $\it{fp}$ shell nuclei in the frame work of large scale shell model calculations. Odd-odd $^{58-62}$Mn isotopes are associated with two shell closures: Z = 20, N = 20 and Z = 20, N = 28. In view of this we have performed two sets of calculations. In the first set valence space is of full $\it{fp}$ shell consisting of 0f$_{7/2}$, 1p$_{3/2}$, 0f$_{5/2}$, 1p$_{1/2}$ orbitals and treating $^{40}$Ca as the inert core. Second set of calculations have been performed in valence space ${fpg_{9/2}}$ taking $^{48}$Ca as inert core. As the dimension of the matrices becomes very large a truncation has been imposed as described ahead.

 For $\it{fp}$ valence space the GXPF1A and KB3G interactions have been used. The configuration space is taken as full $\it{fp}$ shell which is made up of all Pauli allowed combinations of valence particles in the  0f$_{7/2}$, 1p$_{3/2}$, 0f$_{5/2}$ and 1p$_{1/2}$ orbitals for both protons and neutrons. No restriction has been put on the number of particles which can be excited to higher levels. Similar calculations for even isotopes of $^{62-66}$Fe and odd isotopes of $^{61-65}$Fe  were performed by us and have been reported in the literature \cite{Sri09,Sri}. The  ${fpg_{9/2}}$  model space comprises of the  0f$_{7/2}$, 1p$_{3/2}$, 0f$_{5/2}$, 1p$_{1/2}$ active proton orbitals and 0f$_{7/2}$, 1p$_{3/2}$, 0f$_{5/2}$, 1p$_{1/2}$, 0g$_{9/2}$ neutron orbitals with eight 0f$_{7/2}$ frozen neutrons. In ${fpg_{9/2}}$ we used truncation by allowing up to a total of six particle excitations from the 0f$_{7/2}$ orbital to the upper  $\it{fp}$ orbitals for protons and from the upper $\it{fp}$ orbitals to the 0g$_{9/2}$ orbital for neutron as shown in fig.1. The single particle energies for these interactions are given in table 1.
\begin{figure}[ht]
\begin{center}

\resizebox{0.40\textwidth}{!}{
  \includegraphics{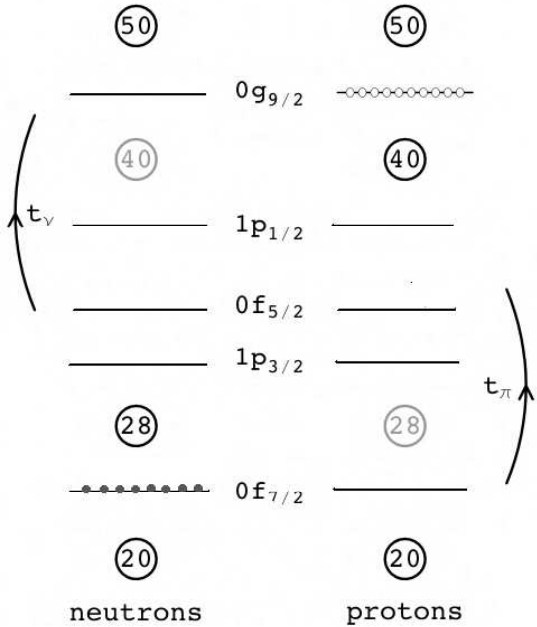}
}
\end{center}
\vspace{-0.3cm}\caption{The truncation procedure used for ${fpg_{9/2}}$ shell in $^{58-62}$Mn isotopes.}
\label{f_turn}
\end{figure}

\begin{table}[ht]
\begin{center}
\caption{
Single particle energies used in GXPF1A, KB3G and $fpg$  interactions in MeV.}
\label{t_spe}
%e\resizebox{7.5cm}{4.5cm}{
\begin{tabular}{rcrc}
\hline
 Orbital  & ~~GXPF1A &~~ KB3G &~~ $fpg$ \cr		   
\hline
 f$_{7/2}$&~~ -8.6240 &~~ -8.6000   &~~ 0.000   \cr
 p$_{3/2}$&~~ -5.6793 &~~ -6.6000   &~~ 2.000   \cr
 p$_{1/2}$&~~ -4.1370 &~~ -4.6000   &~~ 4.000   \cr
 f$_{5/2}$&~~ -1.3829 &~~ -2.1000   &~~ 6.500   \cr
 g$_{9/2}$&~~    -    &~~  -        &~~ 9.000   \cr
\hline              
\end{tabular}
\end{center}
\end{table} 
 
 The GXPF1A interaction is obtained from a fit to the experimental data of unstable nuclei in the $\it{fp}$ shell.  Honma $\it{et~al.}$ \cite{Honma04} have derived an effective interaction, GXPF1, starting from Bonn-C  potential by modifying 70 well determined combinations of four single particle energies and 195 two body matrix elements by iterative fitting calculations  about 699 experimental energy data out of 87 stable nuclei. These authors have tested the GXPF1 interaction for the shell model calculations in the full $\it{fp}$ shell extensively \cite{Honma05} from various viewpoints such as binding energies, electromagnetic moments and transitions, and excitation spectra in the wide range of $\it{fp}$ shell nuclei. They observed that the deviation of the shell model prediction from available experimental data appeared to be  sizable in binding energies of $\it{N}$ $\geq$ 35 nuclei. The $\it{N}$=34 subshell gap in Ca and Ti isotopes is predicted by GXPF1 interaction; however, this shell closure was not observed in recent experimental studies of the $^{52-56}$Ti isotopes \cite{Dinca05b,Fornal04b}, this discrepancy led to the modification of GXPF1 interaction. The interaction was modified by Honma $\it{et~al.}$ by changing five T=1 two body matrix elements in the $\it{fp}$ shell: 3 pairing interaction matrix elements were made slightly weaker and two quadrupole-quadrupole matrix elements were made slightly stronger \cite{Honma05}. The modified interaction, referred to as GXPF1A, gave improved description simultaneously for all these three isotope chains and is reliable for use in shell model calculations to explain the data on unstable nuclei. The KB3G interaction is extracted from KB3 interaction by introducing mass dependence and refining its original monopole changes in order to treat properly the $\it{N=Z}$ = 28 shell closure and its surroundings \cite{Pov01}. 

For the ${fpg_{9/2}}$ valence space, an effective interaction was reported in Ref. \cite{Sorlin02}. This space would generates spurious centre of mass admixtures. Therefore $^{48}$Ca core is considered with 8 frozen neutrons in 0f$_{7/2}$, with  0f$_{7/2}$, 1p$_{3/2}$, 0f$_{5/2}$, 1p$_{1/2}$ active orbitals for protons and 1p$_{3/2}$, 0f$_{5/2}$, 1p$_{1/2}$, 0g$_{9/2}$ active orbitals for neutrons. This interaction was built using $\it{fp}$ two-body matrix elements (TBME) from Ref. \cite{Pov01} and 1p$_{3/2}$, 0f$_{5/2}$, 1p$_{1/2}$, 0g$_{9/2}$ TBME from Ref. \cite{Nowacki96}. For the common active orbitals in these subspaces, matrix elements were taken from Ref. \cite{Nowacki96}. As the latter interaction  was defined for a $^{56}$Ni core, a scaling factor of A$^{-1/3}$ amplitude was applied to take into account the change of radius between the $^{40}$Ca and $^{56}$Ni cores. As protons are added in the $f_{7/2}$ shell, the excitation energy of the 9/2$^+$ state gets decreased due to the attractive interaction $\pi f_{7/2}\nu g_{9/2}$ . The remaining matrix elements were taken from $f_{7/2}$$g_{9/2}$  TBME from Ref. \cite{Kahana69}.

\begin{figure*}[ht]
\begin{center}
\resizebox{0.9\textwidth}{!}{
  \includegraphics{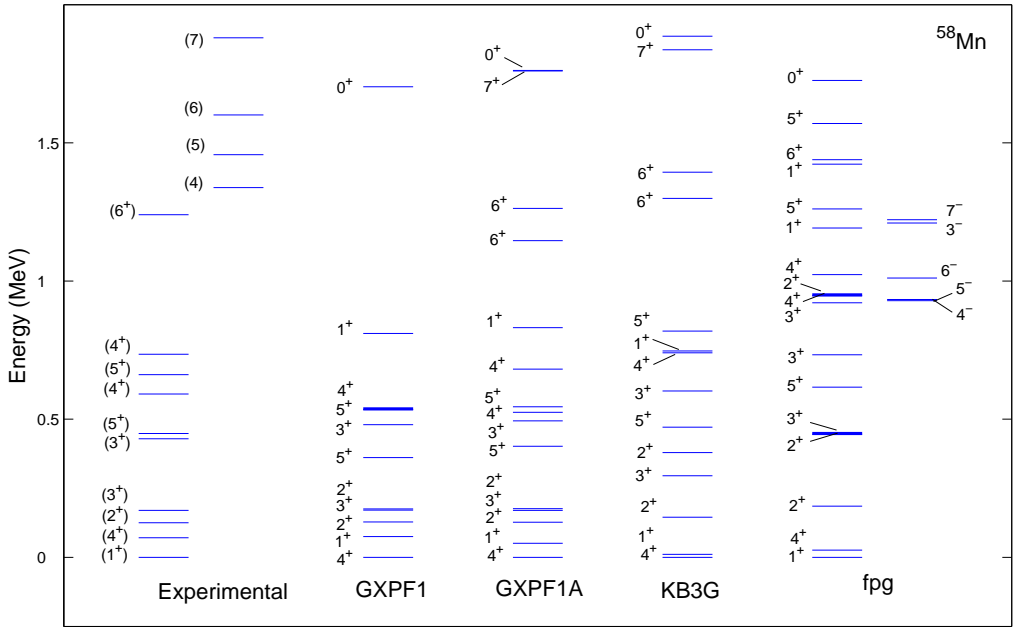}
}
\end{center}
\vspace{-0.3cm}\caption{Comparison of the large-scale shell model calculations of $^{58}$Mn using various effective interaction with the experimental values taken from Refs. \cite{Ste10,NNDC}. The results of GXPF1 interaction taken from Ref. \cite{Liddick06}. }
\label{f_mn_58}
\end{figure*}

  The calculations have been performed  at the SGI-cluster computer at GANIL with the code {\tt ANTOINE }~\cite{Caurier89,Caurier99} and some of the calculations are also performed on the 20 node cluster computer at PRL. In this code the problem of giant matrices is solved by splitting the valence space into two parts, one for proton and another for the neutron. The states of the basis are written as the product of two Slater determinants (SD), one for protons and another for neutrons: $\mid I\rangle = \mid i,\alpha \rangle $ (Here capital letter refers to full space and lower case letters refer to subspaces of proton and neutron).
The Slater determinants i and $\alpha$ can be classified by their $\it{M}$ values, $\it{M_1}$ and $\it{M_2}$. The total $\it{M}$ being fixed, the SD of the two subspaces will be associated only if $\it{M_1+M_2= M}$.

\section{ Results and Discussion}
\subsection{$^{58}$Mn} 
In table 2 all the experimental and theoretical low-lying states up to 2 MeV for $^{58}$Mn are listed. In fig. 2 the calculated energy levels obtained with three different interactions together with the experimental data are shown. The results of earlier calculations \cite{Liddick06} with GXPF1 for comparison are also shown. GXPF1A and KB3G interactions predict the ground state spin and parity as 4$^+$ and the first excited 1$^+$ state at 51 keV and 11 keV respectively. On the other hand $fpg$ interaction predicts the ground state spin 1$^+$ as determine experimentally. The number of predicted 1$^+$ levels below 2 MeV is two for GXPF1A, two for KB3G and three for $fpg$ interaction.   As seen, first 2$^+$ is well predicted by GXPF1A and KB3G interaction while  $fpg$ interaction predicts it 60 KeV higher than experimental value. 
The first 3$^+$ is predicted at 170 keV by GXPF1A, 295 keV by KB3G and 448 keV by $fpg$  interaction while corresponding experimental value is 170 keV.
The first 4$^+$ excited state is predicted at about 26 keV against the experimental value of 71 keV by $fpg$ interaction. It is observed that in going from GXPF1 to GXPF1A and then to KB3G the predicted energy difference between the first excited 1$^+$ state and the 4$^+$ ground state decreases from 75 to 51 and then to 11 keV.   
The experimental data predicts 5$_1$$^+$  level at 448 keV.  In our calculation a 5$_1$$^+$ level is predicted at 402 keV for GXPF1A, 471 keV for KB3G and 616 keV for $fpg$. The 6$_1$$^+$ is also very well predicted by GXPF1A and KB3G interactions.

\begin{figure*}[ht]
\begin{center}

\resizebox{0.90\textwidth}{!}{
  \includegraphics{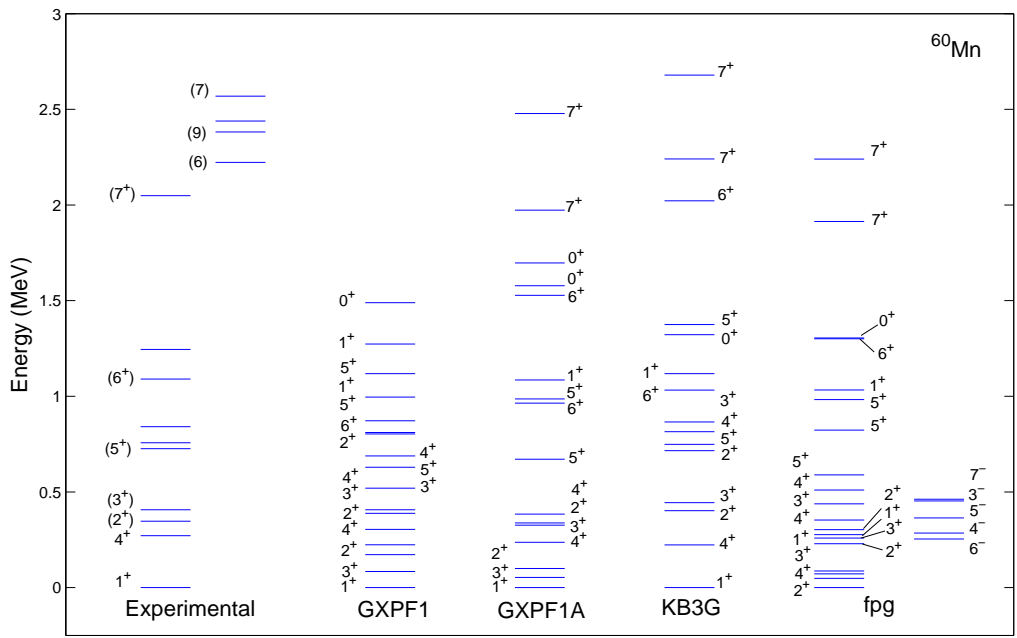}
}
\end{center}
\vspace{-0.3cm}\caption{Comparison of the large-scale shell model calculations of $^{60}$Mn using various effective interaction with the experimental values taken from Refs. \cite{Ste10,NNDC}. The results of GXPF1 interaction taken from Ref. \cite{Liddick06}. }
\label{f_mn_60}
\end{figure*}

 In order to study the importance of the strength (i.e. two-body matrix elements) of the $g_{9/2}$ orbital, we have performed three sets of calculation by allowing 2, 4 and 6 neutrons excitation from 1p$_{3/2}$, 0f$_{5/2}$, 1p$_{1/2}$ to 0$g_{9/2}$.
For 2 neutrons excitation 4$^+$ is the ground state, with a 1$^+$ at 4 keV and 2$^+$ at 182 keV. For  4 neutrons excitation there is a flip of 28 keV giving 1$^+$ as ground state, with a 4$^+$ at 24 keV and 2$^+$ at 186 keV. Finally for 6 neutrons excitation the ground state remains to be the 1$^+$ and 4$^+$, 2$^+$ are at 26 and 186 keV respectively.  But the other low-lying states remain almost same up to 1 MeV (within 200 keV difference). The results for KB3G interaction are close to those from the $fpg$  interaction because $\it{fp}$ two-body matrix elements (TBME) of $fpg$  interaction is taken from KB3G interaction.  

 In more recent experiment, D. Steppenbeck {\it et al.} \cite{Ste10} have reported  many high spin states for $^{58}$Mn above 1338 keV.  In this experiment they observed four levels within 500 keV starting from 1338 keV (these levels are shown in fig. 2), and they make plausible arguments that these levels may be negative parity states. This is an indication of involvement of at least one $g_{9/2}$ neutron. In view of this we have also performed calculations for negative parity states.  In fig. 2 we have shown the results of negative parity states for excitation up to 6 particles into  0$g_{9/2}$ orbital. It is observed that within 300 keV there are four negative parity states with 4$^-$ lying at 929 keV. These levels are more compressed in comparison to experimental findings, which reflects inadequacy of used $fpg$  interaction. Thus for low-lying states up to 1 MeV, KB3G  and GXPF1A both are good. Above 1 MeV, both for high spin positive  and negative parity states it is necessary to include 0$g_{9/2}$ orbital but the results of present $fpg$ interaction indicates that further modification of its present form is required.

\begin{table}
\begin{center}
\caption{
Experimental and theoretical low-lying states (up to 2 MeV) of $^{58}$Mn. Energies are in MeV. The results of GXPF1 interaction are taken from Ref. \cite{Liddick06}.}
\label{t_mn58}
\resizebox{7.0cm}{3.5cm}{
\begin{tabular}{rcrcrcrcrc}
\hline
 J$^\pi$ & ~~Exp. &~~ J$^\pi$ &~~ GXPF1 &~~ J$^\pi$ &~~ GXPF1A &~~  J$^\pi$ &~~ KB3G &~~ J$^\pi$ &~~ $fpg$  \\		   
\hline
(1$^+$)&~~ 0.000 &~~ 4$^+$  &~~ 0.000  &~~ 4$^+$  &~~ 0.000 &~~ 4$^+$  &~~ 0.000 &~~ 1$^+$  &~~ 0.000 \\
(4$^+$)&~~ 0.071 &~~ 1$^+$  &~~ 0.075  &~~ 1$^+$  &~~ 0.051 &~~ 1$^+$  &~~ 0.011 &~~ 4$^+$  &~~  0.026\\
(2$^+$)&~~ 0.125 &~~ 2$^+$  &~~ 0.128  &~~ 2$^+$  &~~ 0.127 &~~ 2$^+$  &~~ 0.145 &~~ 2$^+$  &~~  0.186\\
(3$^+$)&~~ 0.170 &~~ 3$^+$  &~~ 0.171  &~~ 3$^+$  &~~ 0.170 &~~ 3$^+$  &~~ 0.295 &~~ 2$^+$  &~~  0.446 \\
(3$^+$)&~~ 0.429 &~~ 2$^+$  &~~ 0.175  &~~ 2$^+$  &~~ 0.176 &~~ 2$^+$  &~~ 0.379 &~~ 3$^+$  &~~  0.448\\ 
(5$^+$)&~~ 0.448 &~~ 5$^+$  &~~ 0.361  &~~ 5$^+$  &~~ 0.402 &~~ 5$^+$  &~~ 0.471 &~~ 5$^+$  &~~   0.616\\ 
(4$^+$)&~~ 0.591 &~~ 3$^+$  &~~ 0.480  &~~ 3$^+$  &~~ 0.494 &~~ 3$^+$  &~~ 0.602 &~~ 3$^+$  &~~  0.733 \\ 
(5$^+$)&~~ 0.661 &~~ 5$^+$  &~~ 0.535  &~~ 4$^+$  &~~ 0.525 &~~ 4$^+$  &~~ 0.740 &~~ 3$^+$  &~~  0.921\\ 
(4$^+$)&~~ 0.735 &~~ 4$^+$  &~~ 0.537  &~~ 5$^+$  &~~ 0.545 &~~ 1$^+$  &~~ 0.747 &~~ 4$^+$  &~~  0.947\\ 
(6$^+$)&~~ 1.240 &~~ 1$^+$  &~~ 0.810  &~~ 4$^+$  &~~ 0.681 &~~ 5$^+$  &~~ 0.819 &~~ 2$^+$  &~~  0.950 \\
(4)    &~~ 1.338 &~~ 0$^+$  &~~ 1.703  &~~ 1$^+$  &~~ 0.831 &~~ 6$^+$  &~~ 1.299 &~~ 4$^+$  &~~  1.023\\
(5)    &~~ 1.457 &~~  -     &~~   -    &~~ 6$^+$  &~~ 1.146 &~~ 6$^+$  &~~ 1.394 &~~ 1$^+$  &~~  1.192\\ 
(6)    &~~ 1.601 &~~  -     &~~   -    &~~ 6$^+$  &~~ 1.263 &~~ 7$^+$  &~~ 1.837 &~~ 5$^+$  &~~ 1.261\\ 
(7)    &~~ 1.880 &~~  -     &~~   -    &~~ 7$^+$  &~~ 1.760 &~~ 0$^+$  &~~ 1.886&~~ 1$^+$   &~~  1.423\\ 
   -   &~~  -    &~~   -    &~~   -    &~~ 0$^+$  &~~ 1.762 &~~   -    &~~   -  &~~ 6$^+$   &~~  1.439\\ 
   -   &~~  -    &~~   -   &~~    -    &~~   -   &~~    -   &~~   -    &~~   -  &~~ 5$^+$   &~~  1.570\\
   -   &~~  -    &~~   -   &~~    -     &~~   -   &~~    -  &~~   -    &~~   -  &~~ 0$^+$   &~~  1.726\\   
\hline            
\end{tabular}}
\end{center}
\end{table} 

\subsection{$^{60}$Mn}
 In table 3 all the experimental and theoretical low-lying states up to 2.7 MeV for $^{60}$Mn are listed. In fig. 3 the calculated energy levels obtained with three different interactions together with the experimental data are shown. The results of earlier calculations \cite{Liddick06} with GXPF1 for comparison are also shown.  The ground state spin and parity 1$^+$ is correctly reproduced by GXPF1A and KB3G interaction. It is observed that the $fpg$ interaction does not predict correct ground state spin. The calculated 4$^+$ state is at 237 keV for GXPF1A and at 223 keV for KB3G compared to the experimental value of 271 keV. Two states with spin 2$^+$ and 3$^+$ have been predicted to lie between 1$^+$ and 4$^+$ state for GXPF1A. For KB3G these states lie higher than 2$^+$.  The experimental data predicts 5$_1$$^+$  level at 726 keV.  In our calculation a 5$_1$$^+$ level is predicted at 671 keV for GXPF1A, 749 keV for KB3G and 589 keV for $fpg$. The 7$_1$$^+$ level is predicted at 1973 keV for GXPF1A, 2241 keV for KB3G and 1914 keV for $fpg$, however an experimental indication of 7$_1$$^+$  is at 2049 keV. In order to study the importance of $g_{9/2}$ orbital, we have performed three sets of calculation by allowing 2, 4 and 6 neutrons excitation from 1p$_{3/2}$, 0f$_{5/2}$, 1p$_{1/2}$ to 0$g_{9/2}$. The ground state spin and parity is predicted to be 2$^+$ in each case and the other low-lying states remain almost same up to 1 MeV. 

D. Steppenbeck {\it et al.} \cite{Ste10} have reported  many high spin states for $^{60}$Mn above 2223 keV. In this experiment they observed four levels within 350 keV starting from 2223 keV (these levels are shown in fig. 3), and make plausible arguments that these levels may be negative parity states. In fig. 3 we have also shown the results of negative parity states for excitation up to 6 particles into  $g_{9/2}$ orbital.  It is observed that within 208 keV there are 5 negative parity states with 6$^-$ lying at 254 keV. The $fpg$ interaction predicts more compressed levels for negative parity states and also they lie very low in energy. Thus for low-lying states up to 1 MeV, KB3G interaction is far better than GXPF1A interaction. Above 1 MeV, both for high spin positive  and negative parity states it is necessary to include $g_{9/2}$ orbital but the results of present $fpg$ interaction indicates that further modification is required. 

\begin{table}
\begin{center}
\caption{
Experimental and theoretical low-lying states (up to 2.7 MeV) of $^{60}$Mn. Energies are in MeV. The results of GXPF1 interaction taken from Ref. \cite{Liddick06}.}
\label{t_mn60}
\resizebox{8.0cm}{3.5cm}{
\begin{tabular}{rcrcrcrcrc}
\hline
 J$^\pi$ & ~~Exp. &~~ J$^\pi$ &~~ GXPF1  &~~ J$^\pi$ &~~ GXPF1A &~~  J$^\pi$ &~~ KB3G &~~ J$^\pi$ &~~ $fpg$  \\		   
\hline
 1$^+$ &~~ 0.000 &~~ 1$^+$  &~~ 0.000 &~~ 1$^+$  &~~ 0.000 &~~ 1$^+$  &~~ 0.000 &~~ 2$^+$  &~~  0.000\\
 4$^+$ &~~ 0.271 &~~ 3$^+$  &~~ 0.084 &~~ 3$^+$  &~~ 0.053 &~~ 4$^+$  &~~ 0.223 &~~ 4$^+$  &~~  0.048\\
(2$^+$)&~~ 0.347 &~~ 2$^+$  &~~ 0.172 &~~ 2$^+$  &~~ 0.100 &~~ 2$^+$  &~~ 0.402 &~~ 3$^+$  &~~  0.072\\
(3$^+$)&~~ 0.407 &~~ 4$^+$  &~~ 0.224 &~~ 4$^+$  &~~ 0.237 &~~ 3$^+$  &~~ 0.444 &~~ 1$^+$  &~~  0.087\\
(5$^+$)&~~ 0.726 &~~ 2$^+$  &~~ 0.304 &~~ 3$^+$  &~~ 0.326 &~~ 2$^+$  &~~ 0.716 &~~ 2$^+$  &~~  0.229\\ 
   -   &~~ 0.757 &~~ 3$^+$  &~~ 0.388 &~~ 2$^+$  &~~ 0.338 &~~ 5$^+$  &~~ 0.749 &~~ 3$^+$  &~~  0.259\\ 
   -   &~~ 0.841 &~~ 4$^+$  &~~ 0.407 &~~ 4$^+$  &~~ 0.384 &~~ 4$^+$  &~~ 0.815 &~~ 1$^+$  &~~  0.277\\ 
(6$^+$)&~~ 1.090 &~~ 3$^+$  &~~ 0.520 &~~ 5$^+$  &~~ 0.671 &~~ 3$^+$  &~~ 0.866 &~~ 2$^+$  &~~  0.303\\ 
   -   &~~ 1.245 &~~ 5$^+$  &~~ 0.629 &~~ 6$^+$  &~~ 0.964 &~~ 6$^+$  &~~ 1.032 &~~ 4$^+$  &~~  0.353\\ 
(7$^+$)&~~ 2.049 &~~ 4$^+$  &~~ 0.688 &~~ 5$^+$  &~~ 0.986 &~~ 1$^+$  &~~ 1.118 &~~ 3$^+$  &~~  0.438\\
 (6)   &~~ 2.223 &~~ 2$^+$  &~~ 0.803 &~~ 1$^+$  &~~ 1.085 &~~ 0$^+$  &~~ 1.322 &~~ 4$^+$  &~~  0.510\\
 (9)   &~~ 2.382 &~~ 6$^+$  &~~ 0.811 &~~ 6$^+$  &~~ 1.528 &~~ 5$^+$  &~~ 1.375 &~~ 5$^+$  &~~  0.589\\ 
   -   &~~ 2.439 &~~ 5$^+$  &~~ 0.872 &~~ 0$^+$  &~~ 1.578 &~~ 6$^+$  &~~ 2.022 &~~ 5$^+$  &~~  0.823\\ 
 (7)   &~~ 2.569 &~~ 1$^+$  &~~ 0.996 &~~ 0$^+$  &~~ 1.697 &~~ 7$^+$  &~~ 2.241 &~~ 5$^+$  &~~  0.983\\ 
   -   &~~   -   &~~ 5$^+$  &~~ 1.118 &~~ 7$^+$  &~~ 1.973 &~~ 7$^+$  &~~ 2.679 &~~ 1$^+$  &~~  1.033\\ 
   -   &~~   -   &~~ 1$^+$  &~~ 1.273 &~~ 7$^+$  &~~ 2.478 &~~   -    &~~   -   &~~ 6$^+$  &~~  1.301\\
   -   &~~   -   &~~ 0$^+$  &~~ 1.489 &~~   -    &~~   -   &~~   -    &~~   -   &~~ 0$^+$  &~~  1.305\\
   -   &~~   -   &~~        &~~       &~~   -    &~~   -   &~~   -    &~~   -   &~~ 7$^+$  &~~  1.914\\
   -   &~~   -   &~~        &~~       &~~   -    &~~   -   &~~   -    &~~   -   &~~ 7$^+$  &~~  2.240\\ 
\hline             
\end{tabular}}
\end{center}
\end{table} 

\begin{figure}[ht]
\begin{center}
\resizebox{0.5\textwidth}{!}{
  \includegraphics{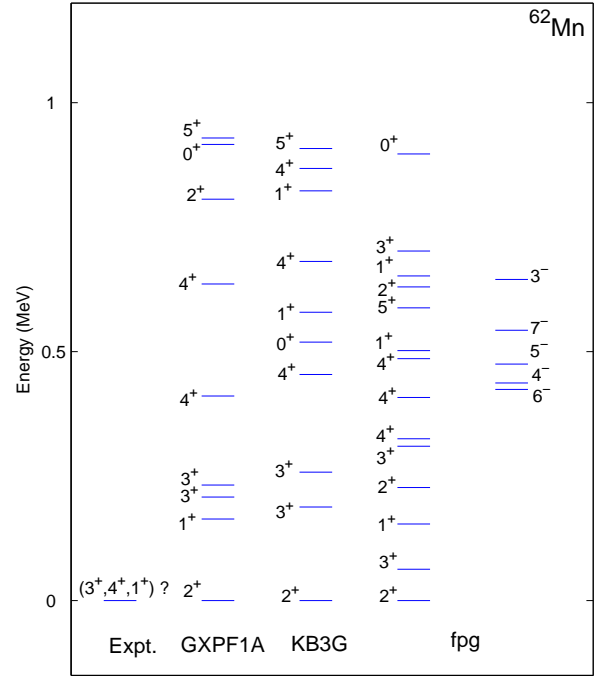}
}
\end{center}
\vspace{-0.3cm}\caption{Comparison of the large-scale shell model calculations of $^{62}$Mn using various effective interaction with the experimental values taken from Refs. \cite{Gau05,NNDC}.}
\label{f-mn-62}
\end{figure}

\subsection{$^{62}$Mn}

In table 4 we list experimental ground state and all theoretical low-lying states up to 1 MeV for $^{62}$Mn isotope. In fig. 4 we have shown the comparison of experimental values with those predicted for three different interactions.
 Experimentally for $^{62}$Mn there is an uncertainty in assigning the ground state spin to be 1$^+$, 3$^+$ or 4$^+$. All the calculated results show 2$^+$ as a ground state. In order to study the importance of $g_{9/2}$ orbital for this neutron rich nuclei, we have performed three sets of calculation by allowing 2, 4 and 6 neutrons excitation from 1p$_{3/2}$, 0f$_{5/2}$, 1p$_{1/2}$ to 0$g_{9/2}$. The ground state spin and parity is predicted to be 2$^+$ in each case and the other low-lying states remain almost same up to 1 MeV. In fig. 4 we have also shown the results of negative parity states for six particle excitation. It is observed that within 221 keV there are 5 negative parity states with 6$^-$ lying at 424 keV. The predicted results of the $fpg$  interaction may be relevant. However the experimental data is very sparse, thus it is not possible to make any definite conclusions regarding the prediction of the $fpg$  interaction.

\begin{table}
\begin{center}
\caption{
Experimental and theoretical low-lying states (up to 1 MeV) of $^{62}$Mn. Energies are in MeV.}
\label{t_mn62}
\resizebox{7.5cm}{3.5cm}{
\begin{tabular}{rcrcrcrc}
\hline
 J$^\pi$ & ~~Exp. &~~ J$^\pi$ &~~ GXPF1A &~~  J$^\pi$ &~~ KB3G &~~ J$^\pi$ &~~ $fpg$ \\		   
\hline
(3$^+$,4$^+$,1$^+$)? &~~ 0.000 &~~ 2$^+$  &~~ 0.000 &~~ 2$^+$ &~~ 0.000&~~ 2$^+$ &~~  0.000  \\
   -   &~~  -    &~~ 1$^+$  &~~ 0.164 &~~ 3$^+$  &~~ 0.188&~~ 3$^+$  &~~  0.063 \\
   -   &~~  -    &~~ 3$^+$  &~~ 0.208 &~~ 3$^+$  &~~ 0.258&~~ 1$^+$  &~~  0.154  \\
   -   &~~  -    &~~ 3$^+$  &~~ 0.232 &~~ 4$^+$  &~~ 0.454&~~ 2$^+$  &~~  0.227  \\
   -   &~~   -   &~~ 4$^+$  &~~ 0.411 &~~ 0$^+$  &~~ 0.519&~~ 3$^+$  &~~  0.310\\ 
   -   &~~   -   &~~ 4$^+$  &~~ 0.636 &~~ 1$^+$  &~~ 0.579&~~ 4$^+$  &~~   0.325\\ 
   -   &~~   -   &~~ 2$^+$  &~~ 0.806 &~~ 4$^+$  &~~ 0.681&~~ 4$^+$  &~~  0.408 \\ 
   -   &~~   -   &~~ 0$^+$  &~~ 0.916 &~~ 1$^+$  &~~ 0.823&~~ 4$^+$  &~~  0.486\\ 
  -    &~~   -   &~~ 5$^+$  &~~ 0.929 &~~ 4$^+$  &~~ 0.868&~~ 1$^+$  &~~  0.502 \\ 
   -   &~~   -   &~~   -    &~~  -    &~~ 5$^+$  &~~ 0.908&~~ 5$^+$  &~~  0.588\\ 
  -   &~~   -   &~~   -    &~~  -     &~~  -     &~~   -  &~~ 2$^+$  &~~  0.630\\ 
  -   &~~   -   &~~   -    &~~  -     &~~  -     &~~   -  &~~ 1$^+$  &~~  0.652\\ 
  -   &~~   -   &~~   -    &~~  -     &~~  -      &~~  -  &~~ 3$^+$  &~~  0.702\\   
  -   &~~   -   &~~   -    &~~  -     &~~  -      &~~  -  &~~ 0$^+$  &~~  0.897\\   
\hline           
\end{tabular}}
\end{center}
\end{table}

\begin{table}
\caption{The extent of configuration mixing involved in $^{58-62}$Mn
isotopes for different states. For each state the numbers quoted are $S$,  sum  of  contributions  from  particle
partitions  each  of  which  is contributing greater than 1\%; $M$, maximum contribution from  a  single  partition, and
$N$, total number of partitions contributing in $S$.}
\begin{center}
\begin{tabular}{cccc}
\hline
J$^\pi$&$^{58}_{25}$Mn$_{33}$&J$^\pi$& $^{60}_{25}$Mn$_{35}$\cr

       &{$S$, $M$, $N$}   &         &{$S$, $M$, $N$}        \cr
\hline
GXPF1A\\
4$_{gs}^+$&~74.7,~25.9,~16& 1$_{gs}^+$&~~82.9,~34.7,~16\cr
1$_{1}^+$&~73.1,~22.2,~16&  3$_{1}^+$&~~84.5,~39.6,~16 \cr
5$_{1}^+$&~76.1,~22.0,~17&  5$_{1}^+$&~~84.5,~23.0,~17\cr\\
KB3G\\
4$_{gs}^+$&~79.0,~35.4,~17& 1$_{gs}^+$&~~82.3,~42.3,~12\cr
1$_{1}^+$&~76.3,~28.1,~16&  4$_{1}^+$&~~82.9,~36.5,~13\cr
5$_{1}^+$&~80.3,~36.4,~19&  5$_{1}^+$&~~84.4,~35.4,~15\cr\\
\hline
J$^\pi$&$^{62}_{25}$Mn$_{37}$\\
        &{$S$, $M$, $N$}\\
\hline
$fpg$ \\

 2$_{gs}^+$&~69.5,~13.8,~21\cr
 3$_{1}^+$&~72.0,~14.1,~21\cr
 5$_{1}^+$&~72.0,~15.6,~20\cr\\

\hline
\end{tabular}
\end{center}
\end{table}

\subsection{Wave functions}

Structure of the wave functions for the ground state and two excited states 
of $^{58-62}$Mn for the three different interactions are compared in table 5. 
In this table we have tabulated (i) $S$, the sum of contribution from particle partitions having
 contribution greater than 1\%. 
(ii) $M$, the maximum contribution from a single partition, and
(iii)  $N$, the total number of partitions contributing to  $S$. 
The deviation of  $S$ from 100\% is due to high configuration mixing.
The increase in  $N$ is also a signature of larger
configuration mixing. From the table one can see that for $^{58,60}$Mn, $S$ is changing from $\sim$ 73\% to $\sim$ 84\%; the number of partition is changing from 12
- 19 and $M$ is changing from $\sim$ 22\% to $\sim$ 42\%. For $^{58}$Mn, 4$^+$ is the ground state predicted by GXPF1A 
and KB3G interactions with $\pi(0f^5_{7/2})$$\otimes$$\nu(0f^8_{7/2}1p^3_{3/2}0f^2_{5/2})$ partition having intensity 25.9\% and 35.4\%  respectively. The first 5$^+$ has $\pi(0f^5_{7/2}$)$\otimes$$\nu$$(0f^8_{7/2}1p^3_{3/2}0f^2_{5/2})$ partition with intensity 22.0\% and 36.4\% for GXPF1A and KB3G respectively. For $^{60}$Mn, KB3G interaction predicts better spectra and for this interaction $\pi(0f^5_{7/2})$$\otimes$$\nu$$(0f^8_{7/2}1p^4_{3/2}0f^3_{5/2})$ partition has 42.3\% intensity for the ground state.
For $^{62}$Mn, the value of $S$ is $\sim$ 70\%; $N$ is 21 and $M$ is $\sim$ 14\%  for the 2$^+$ ground state.
The ground state has $\pi(0f^5_{7/2})$$\otimes$$\nu$($0f^8_{7/2}1p^4_{3/2}\\0f^2_{5/2}
1p^1_{1/2} 0g^2_{9/2})$ partition with intensity 13.8\%. The occupancy of $g_{9/2}$ orbital for the ground state is 2.09.

\section{Summary}

In the present work large scale shell model calculations have been performed for neutron rich odd-odd isotopes of Mn with A=58, 60, 62 in two valence spaces: full $\it {fp}$ space and ${fpg_{9/2}}$ space using $^{48}$Ca as core. For $\it {fp}$ space, calculations have been performed with recently derived GXPF1A and KB3G interactions without a truncation. For  ${fpg_{9/2}}$ space, calculations have been performed with $fpg$ interaction using a truncation. In going from $^{58}$Mn to $^{60}$Mn the experimental value for the separation between first excited 4$^+$ state and 1$^+$ ground state increases from 71 keV to 271 keV. Even though the $fpg$  interaction in a truncated space does not reproduce ground state spin for $^{60}$Mn it is likely that expanding the valence space to include more orbitals from $\it{sdg}$ shell and/or more tuning of $fpg$  interaction is required.  For low-lying states in $^{58}$Mn the KB3G and GXPF1A both predicts good results and for $^{60}$Mn, KB3G is better than GXPF1A. For negative parity and high-spin positive parity states for both isotopes, a good interaction in ${fpg_{9/2}}$ space is essential. For $^{62}$Mn, the predicted results of $fpg$  interaction may be relevant. However the experimental data is very sparse, thus it is not possible to make any definite conclusions regarding the prediction of the $fpg$  interaction. Presently we are attempting to develop a suitable $fpg$ interaction for neutron rich $fp$ shell nuclei. A key feature yet to be identified in the neutron rich odd-odd Mn isotopes is the location of the negative parity levels that could signify the presence of the 0$g_{9/2}$ orbital. More experimental data on higher neutron-rich Mn isotopes are needed to discuss the onset of the collectivity while approaching $\it{N = }$40.\\ 

 The first author would like to thank P.~Van Isacker, M.~Rejumund, E.~Caurier and F.~Nowacki for providing access to the shell model code ANTOINE at GANIL. We are grateful to V.K.B.~Kota for his interest and useful discussions. This work was financially supported by the Sandwich PhD programme of the Embassy of France in India.

\end{fmffile}
\end{document}